\documentstyle[prd,aps,floats,epsf]{revtex}

\newcommand{\StruveL}{{\mbox{\bf L}}}
\newcommand{\StruveH}{{\mbox{\bf H}}}

\def\spose#1{\hbox to 0pt{#1\hss}}
\def\ltapprox{\mathrel{\spose{\lower 3pt\hbox{$\mathchar"218$}}
 \raise 2.0pt\hbox{$\mathchar"13C$}}}
\def\gtapprox{\mathrel{\spose{\lower 3pt\hbox{$\mathchar"218$}}
 \raise 2.0pt\hbox{$\mathchar"13E$}}}
\def\inapprox{\mathrel{\spose{\lower 3pt\hbox{$\mathchar"218$}}
 \raise 2.0pt\hbox{$\mathchar"232$}}}

\begin{document}

\title{\Large{{\bf Universal Scaling of the Chiral Condensate \\
in Finite-Volume Gauge Theories}}}

\vspace{1.5cm}

\author{~\\~\\
{\sc Poul H. Damgaard}\\
The Niels Bohr Institute\\ Blegdamsvej 17\\ DK-2100 Copenhagen\\
Denmark\\~\\
{\sc Robert G. Edwards, Urs M. Heller}\\
Supercomputer Computations Research Institute\\
Florida State University\\Tallahassee, FL 32306-4130\\USA\\~ \\and\\~ \\
{\sc Rajamani Narayanan}\\American Physical Society\\
One Research Road\\ Ridge, NY 11961\\USA}
\maketitle
\vfill
\begin{abstract} 
We confront exact analytical predictions for the finite-volume
scaling of the chiral condensate with data from quenched lattice gauge 
theory simulations. Using staggered fermions in both the
fundamental and adjoint representations, and gauge groups SU(2) 
and SU(3), we are able to test simultaneously all of the three 
chiral universality classes. With overlap fermions we also
test the predictions for gauge field sectors of non-zero topological 
charge. Excellent agreement is found in most cases, and the deviations
are understood in the others. 
\end{abstract}
\vfill
\begin{flushleft}
NBI-HE-99-23 \\
FSU-SCRI-99-41\\
hep-lat/9907016
\end{flushleft}
\thispagestyle{empty}
\newpage

\setcounter{page}{1}
\section{Introduction}

The constraints imposed by chiral symmetry breaking in gauge theories
can be surprisingly strong. Low-energy theorems, the dynamics of 
pseudo-Goldstone bosons in an expansion around the zero-momentum limit,
and the whole framework of effective chiral Lagrangians are examples
of this. Generally these constraints are imposed on the effective 
low-energy degrees of freedom only. It is much more surprising that
both spontaneous chiral symmetry breaking and the explicit breaking
of chiral symmetry due to the U(1) anomaly also can be used to give exact
analytical predictions for the underlying {\em fermion} degrees of freedom.
This is possible when one restricts the gauge theory to a large but finite
four-volume $V$ obeying the inequality \cite{LS}
\begin{equation}
 V ~<<~ \frac{1}{m_{\pi}^4} ~,
\end{equation}
where $m_{\pi}$ is the pseudo-Goldstone mass. In this rather extreme
limit the QCD partition function depends on the fermion masses $m_i$ only 
in the particular combination $\mu_i \equiv V\Sigma m_i$, where $\Sigma$ 
is the chiral condensate. While the four-volume $V$ must be taken to infinity
in order to obtain analytical predictions, a finite-size scaling regime
is thus achieved by sending fermion masses $m_i$ to zero at just such a rate 
that the $\mu_i$'s remain fixed. This is an exact finite-size scaling region
in the same sense as near critical points: we can reach as accurate 
agreement we wish by simply choosing a sufficiently large four-volume
$V$. In contrast to what one is accustomed to in statistical mechanics,
the finite-size scaling functions are in this case known {\em exactly}, both in
shape and absolute normalization, once one has the value of the 
infinite-volume chiral condensate $\Sigma$.

One of the remarkable aspects of the finite-size scaling region (1) is 
its relation to Random Matrix Theory results that have proven
to be universal \cite{SV,V,ADMN}.
There is a beautiful classification of the possible chiral symmetry
breaking patterns for different gauge groups and color representations
of the fermions in terms of the classical Random Matrix Theory ensembles
labeled by the so-called Dyson index. This leads to three major
universality classes \cite{SmV}. For the quenched case, the
analytical prediction for the mass-dependent chiral condensate
was in fact first obtained by Verbaarschot \cite{V0} using the exact 
formula for the microscopic Dirac operator spectrum as derived from 
Random Matrix Theory. It has later become clear that these results can also 
be derived from finite-volume partition functions alone \cite{AD}. A powerful
analytical technique uses fully or partially quenched (supersymmetric)
chiral Lagrangians \cite{OTV}. Lattice gauge simulations have already
shown nice agreement with the exact analytical predictions for the
microscopic Dirac operator spectrum associated with all three different
universality classes \cite{BBetal,DHK,EHN} using staggered fermions. It
has been particularly challenging to see also the detailed analytical
predictions in gauge field sectors of fixed non-zero topological charge $\nu$,
and this has very recently been achieved using perfect actions \cite{Lang}
and overlap fermions \cite{EHKN}.

The purpose of this paper is to perform a systematic series of lattice gauge 
theory tests of the exact predictions related to the 
mass-dependent chiral condensates and one of the chiral susceptibilities.
For the case of gauge group SU(3) and staggered fermions in
the fundamental representation such an analysis was first performed 
\cite{V0} on the basis of lattice gauge theory data from the Columbia group
\cite{Columbia}. These data, based on configurations with $N_f=2$ dynamical
fermions, were taken for finite-temperature lattice volumes, and with
rather large correlation lengths. Because the dynamical fermion masses were
large on that scale, and the physical pions therefore not obeying 
the inequality (1), the configurations were in fact to be considered as 
completely 
quenched on the scale of the mass of the valence fermions. As there are 
very accurate estimates for the infinite-volume chiral condensate $\Sigma$ 
for the same theory at different $\beta$-values \cite{DHK}, we now 
have parameter-free predictions at these $\beta$-values. Furthermore, we
can probe much larger physical volumes (and hence obtain higher
accuracy), and our result will not be contaminated by 
finite-temperature effects. This case corresponds to the chiral {\em unitary}
(chUE) Random Matrix Theory ensemble. We next turn to gauge group SU(2) and 
staggered
fermions in the fundamental and adjoint representations. The former case
corresponds at our finite lattice spacings to the chiral {\em symplectic}
ensemble (chSE) in the Random Matrix Theory classification, while the latter 
corresponds to the chiral {\em orthogonal} ensemble (chOE). 
In this way we cover all three major
universality classes. However, staggered fermions suffer from two significant
defects in this context. First, at our finite lattice spacings the staggered
fermions with SU(2) gauge group do not fall into the right universality 
classes as compared with fermions in the continuum \cite{V}. Second, artifacts
due to finite lattice spacings prevent us from testing more than the
gauge field sector of topological charge $\nu=0$ with these fermions. 
Both of these shortcomings
can be overcome by the use of more sophisticated fermion formulations.
We shall here provide lattice data obtained with overlap fermions
\cite{herbert}. 
Here finite lattice-spacing analogs of continuum relations
for the chiral condensate in non-trivial topological backgrounds can be
established \cite{EHN1}. We thus simultaneously achieve both the correct 
identification with respect to continuum universality 
classes, and correct relationships in non-trivial topological gauge field
sectors. We shall throughout restrict ourselves to the quenched limit,
$N_f=0$. Analytically this limit is readily taken, both from
Random Matrix Theory and from the (quenched) finite-volume partition functions,
and the answers have been shown to agree. There are thus precise and
unequivocal analytical predictions also for this case.

For which observables do we have exact analytical predictions? Essentially
all quantities for which the partition function itself ${\cal Z}(\{\mu_i\})$, 
perhaps extended with additional fermion species, is a generating function.
The simplest quantity to focus on is obviously the mass-dependent
chiral condensate itself, which in the quenched theory simply reads
\begin{equation}
\frac{\Sigma(\mu)}{\Sigma} ~\equiv~ 
\frac{\partial}{\partial\mu}\ln[{\cal Z}(\mu)] ~.
\end{equation}
Here $\Sigma$ is the genuine
chiral condensate in the massless limit of the infinite-volume theory, 
and $\mu=V\Sigma m$ is the microscopically rescaled quenched ``valence'' 
fermion mass $m$. Because of the relation (at fixed topological charge $\nu$) 
\begin{equation}
\frac{\Sigma_{\nu}(\mu)}{\Sigma} ~=~ 2\mu\int_0^{\infty}d\zeta
\frac{\rho^{(\nu)}_s(\zeta)}{\zeta^2+\mu^2} + \frac{|\nu|}{\mu}~,
\label{chirho}
\end{equation}
the mass-dependent chiral condensate tests a massive spectral sum rule
for the microscopic density $\rho_s^{(\nu)}(\zeta;\mu)$ of the Dirac operator 
spectrum \cite{SV,masssum}.\footnote{In what follows we shall for
convenience, to avoid absolute-value signs, 
always consider $\nu$ non-negative.} We shall here restrict ourselves to
the quenched cases, where $\rho_s^{(\nu)}(\zeta)$ is mass-independent
(and the partially quenched cases are completely analogous).
This is one way in which the analytical prediction for the
quenched limit can be obtained (the other proceeds directly from the quenched
chiral Lagrangian \cite{OTV}). In numerical simulations the condensate 
measurements are of course performed very easily (from the trace of the 
propagator), without ever having to compute the Dirac operator spectrum 
itself. Nevertheless, we have sometimes found it convenient to supplement
direct measurements by appropriate sums over eigenvalues. In this connection
it is important to stress the following point. Because of Eq.~(\ref{chirho}),
we are in fact concerned with tests of the analytical predictions for 
$\rho_s(\zeta,\mu)$. The advantage of using $\Sigma_{\nu}(\mu)$ to test
these predictions is that it in a quantitative manner probes the 
microscopic spectral density in different regimes. For instance, by
going to very small values of $\mu$ the condensate becomes very sensitive
to the precise eigenvalue distribution around $\zeta \sim \mu$. In particular
for the chUE (in the $\nu=0$ sector) and the chOE (in the $\nu=0$ and
$\nu=1$ sectors) $\Sigma_{\nu}(\mu)$, as we shall see below, is extremely
sensitive to the low-$\zeta$ distribution of the smallest non-zero
eigenvalues. This effect can be enhanced by considering in addition an
observable such as a chiral susceptibility, as we shall discuss below.

Considering the expression (\ref{chirho}) for the chiral condensate,
one might wonder about the necessity of subtractions. After all, even in
free field theory the spectral density of the Dirac operator goes
like $\rho(\lambda) \sim \lambda^3$, and the spectral representation
of the condensate is thus ultraviolet divergent. There are no such
divergences in the finite-volume scaling regime considered here, and we
should make no subtractions in the chiral condensate either. Although
this point was already explained in the paper of Leutwyler and Smilga
\cite{LS}, it is worthwhile to repeat it here. The explanation is as
follows: What we are computing here is not the conventionally defined
chiral condensate. We are taking a correlated limit of $V\to\infty$ and
$m\to 0$ such that $\mu \equiv m\Sigma V$ is kept fixed. In this limit
the condensate $\Sigma(\mu)$ receives contributions only from Dirac operator
eigenvalues on the scale of $m \ll \Lambda_{QCD}$, and below. The ultraviolet
end of the Dirac operator spectrum is not ignored: the corrections
(and hence subtractions) from this region are of the kind $m\Lambda^2$ and 
$m^3\ln\Lambda$, where $\Lambda$ is the ultraviolet cutoff. In the scaling
region where $\mu=m\Sigma V$ is kept fixed, these terms are suppressed
by $1/V$ and $1/V^3$, respectively. In other words, the ultraviolet end of
the Dirac operator spectrum in this region enters only as $1/V$ corrections
to the main predictions. In addition, there are of course also $1/V$
corrections from the neglect of non-static modes in the effective 
partition function,
so all of these $1/V$ corrections are effectively beyond our control.
Consistent with this observation is the fact that $1/V$ corrections are
non-universal in the Random Matrix Theory context \cite{top}. Here we
are interested only in the leading, universal, predictions for $V\to\infty$.

Corresponding to the three different universality classes, there are three
distinct predictions for the mass-dependent condensate. We shall here
give the predictions for gauge field sectors of fixed topological
index $\nu$, and for any number of massless flavors $N_f$. First, for gauge
groups SU($N_c$) with $N_c\geq3$ and fermions in the fundamental
representation the prediction reads \cite{V0}
\begin{equation}
\frac{\Sigma_{\nu}^{\rm{\mbox chUE}}(\mu)}{\Sigma} = 
\mu\left[I_{N_{f}+\nu}(\mu)K_{N_{f}+\nu}(\mu)
+ I_{N_{f}+\nu+1}(\mu)K_{N_{f}+\nu-1}(\mu)\right]  + \frac{\nu}{\mu} ~, 
\label{sigmaUE}
\end{equation}
where $I_n(x)$ and $K_n(x)$ are the two modified Bessel functions. This
is the universality class of the chUE in the Random Matrix Theory
classification. Staggered fermions in the same representation and for
the same gauge groups are here belonging to the same universality class
as continuum fermions. In this case the explicit form of the microscopic
spectral density $\rho_s^{(\nu)}(\zeta;\mu_i)$ of also $N_f$ {\em massive}
fermions (of masses $\mu_i$)
is known, so that one has also available complete analytical
predictions for partially quenched chiral condensates with massive
fermions. These can also be derived directly from partially quenched
chiral Lagrangians (see ref. \cite{OTV}).

For the chSE universality class, where the microscopic spectral density
for $N_f$ massless fermions in a sector of arbitrary topological charge
$\nu$ has been given in very compact form in \cite{BBetal}, 
we have been able to reduce the chiral condensate to the following.
When $N_f$ is {\em even} we find 
\begin{eqnarray}
 \frac{\Sigma_{\nu}^{\rm{\mbox chSE}}(\mu)}{\Sigma} &=&  2\mu\left[
I_{N_{f}+2\nu}(2\mu)K_{N_{f}+2\nu}(2\mu)
+ I_{N_{f}+2\nu+1}(2\mu)K_{N_{f}+2\nu-1}(2\mu)\right]  + \frac{\nu}{\mu}
\cr &-& 2(-1)^{\nu+N_{f}/2}
K_{N_{f}+2\nu}(2\mu)\left[\mu\left(I_{0}(2\mu)
- \frac{\pi}{2}(\StruveL_0(2\mu)I_1(2\mu)-\StruveL_1(2\mu)I_0(2\mu))\right)
\right.\cr
&& \qquad - \left.\sum_{k=0}^{\nu+N_{f}/2-1} (-1)^kI_{2k+1}(2\mu)\right]
\label{sigmaSEeven}
\end{eqnarray}
where $\StruveL_n(x)$ denotes the $n$th order modified Struve function. 
For $N_f$ {\em odd} we find
\begin{eqnarray}
 \frac{\Sigma_{\nu}^{\rm{\mbox chSE}}(\mu)}{\Sigma} &=&  2\mu\left[
I_{N_{f}+2\nu}(2\mu)K_{N_{f}+2\nu}(2\mu)
+ I_{N_{f}+2\nu+1}(2\mu)K_{N_{f}+2\nu-1}(2\mu)\right]  + \frac{\nu}{\mu}
\cr &-& (-1)^{\nu+(N_{f}+1)/2}
K_{N_{f}+2\nu}(2\mu)\left[1 - I_0(2\mu) 
- 2\sum_{k=1}^{\nu+(N_{f}-1)/2} (-1)^k I_{2k}(2\mu)\right]
\label{sigmaSEodd}
\end{eqnarray}
These
predictions pertain to gauge group SU($N_c$) with $N_c\geq2$ and fermions
in the adjoint representation (for continuum fermions). This is also the
universality class relevant for staggered fermions in the fundamental
representation and gauge group SU(2). Finally, the universality class
of the chOE predicts a chiral condensate of the following form. For
$\nu$ {\em odd} we find, using the recent compact expression for
the microscopic spectral density of that case \cite{Widom},
\begin{eqnarray}
 \frac{\Sigma_{\nu}^{\rm{\mbox chOE}}(\mu)}{\Sigma} &=& \mu\left[
I_{2N_{f}+\nu}(\mu)K_{2N_{f}+\nu}(\mu)
+ I_{2N_{f}+\nu+1}(\mu)K_{2N_{f}+\nu-1}(\mu)\right]
 + \frac{\nu}{\mu} \cr
&+& (-1)^{N_{f}+\frac{\nu-1}{2}}\left[I_{2N_{f}+\nu}(\mu)K_0(\mu)
+ 2\sum_{k=1}^{N_{f}+(\nu-1)/2}(-1)^k I_{2N_{f}+\nu}(\mu)K_{2k}(\mu)\right] ~,
\label{sigmaOE1}
\end{eqnarray}
while for $\nu$ {\em even} we have been able to reduce the answer to
\begin{eqnarray}
 \frac{\Sigma_{\nu}^{\rm{\mbox chOE}}(\mu)}{\Sigma} &=&  
\mu\left[I_{2N_{f}+\nu}(\mu)K_{2N_{f}+\nu}(\mu)
+I_{2N_{f}+\nu+1}(\mu)K_{2N_{f}+\nu-1}(\mu)\right] + \frac{\nu}{\mu} \cr
&+&(-1)^{N_{f}+\nu/2}\frac{\pi}{2}\left[I_{2N_{f}+\nu}(\mu)-
\StruveL_{2N_{f}+\nu}(\mu)\right] + \sum_{k=0}^{N_{f}+\nu/2-1}(-1)^k 
\frac{(2N_{f}+\nu-2k-3)!!}{(2N_{f}+\nu+2k+1)!!} \mu^{2k+1} \cr
&-&(-1)^{N_{f}+\nu/2}K_{2N_{f}+\nu}(\mu)\left[\mu\left(I_0(\mu)-
\frac{\pi}{2}(\StruveL_0(\mu)I_1(\mu)-\StruveL_1(\mu)I_0(\mu))\right)\right.
\cr && \qquad -2 \left.\sum_{k=0}^{N_{f}+\nu/2-1}(-1)^k I_{2k+1}(\mu)\right]
\label{sigmaOE2}
\end{eqnarray}
where, with the usual convention, $(-1)!! \equiv 1$.
Eqs.~(\ref{sigmaOE1}) and (\ref{sigmaOE2}) give the chiral 
condensate in SU(2) gauge theory and
continuum fermions in the fundamental (pseudo-real) representation. 
They also correspond to gauge group
SU($N_c$) with $N_c\geq2$ and staggered fermions in the adjoint 
representation.

\begin{figure}
\centerline{{\setlength{\epsfxsize}{6in}\epsfbox[80 40 550 530]{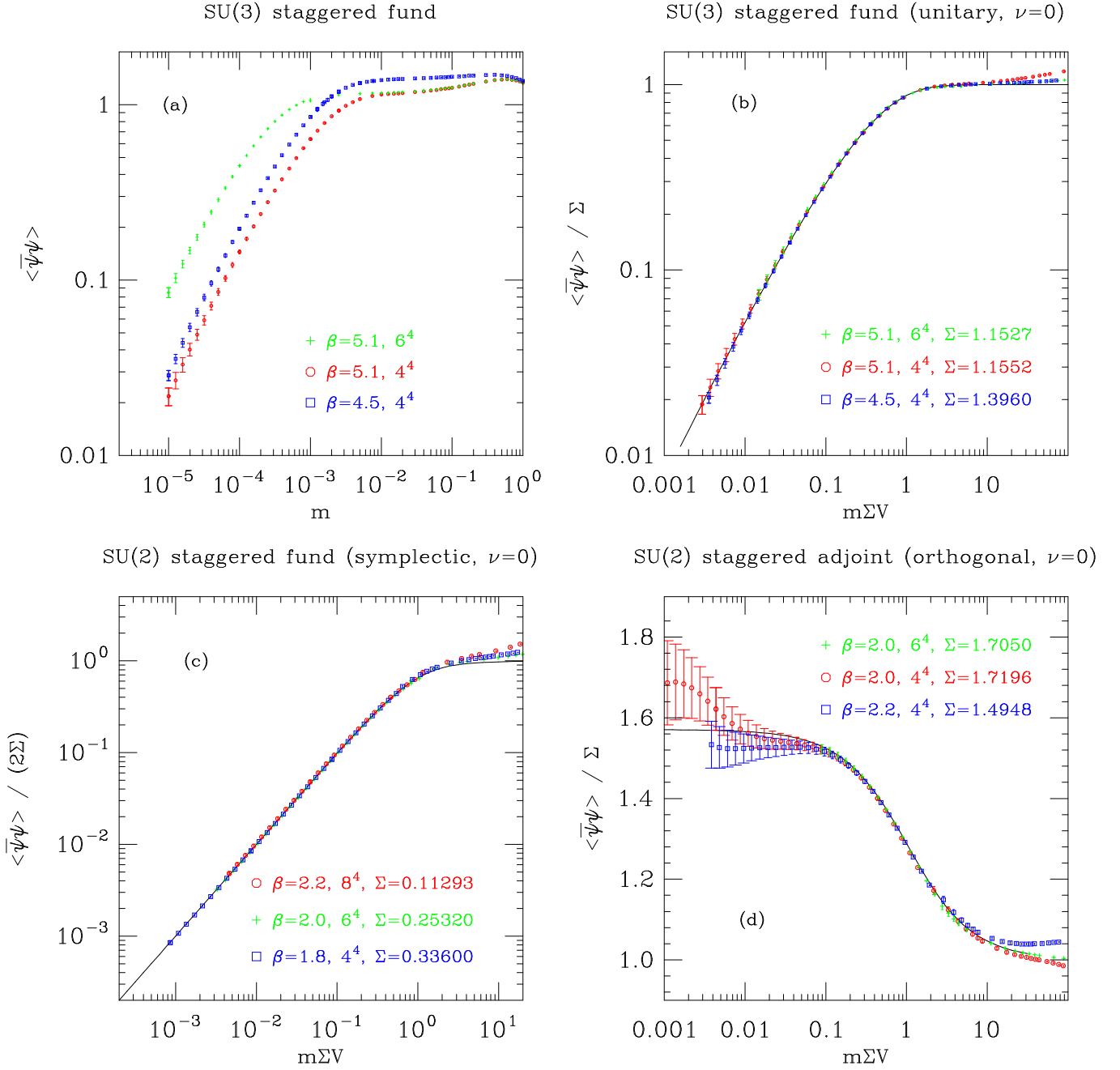}}}
\caption{The condensate for staggered fermions in (a) the fundamental
representation of SU(3) as function of the fermion mass, 
(b) rescaled form as a function of $\mu = m \Sigma V$, 
(c) the fundamental
representation of SU(2),
(d) the adjoint
representation of SU(2).}
\label{fig:st_pbp}
\end{figure}

\section{Staggered Fermions}

Analytical predictions for the quenched chiral condensate and higher
chiral susceptibilities are all restricted to sectors of fixed gauge
field topology.\footnote{With dynamical fermions one can analytically
perform the required sum over topology \cite{LS}, but this is not
possible in the quenched case without additional assumptions about the 
distribution of winding numbers \cite{top}.} As mentioned above, 
with staggered fermions lattice
simulations of the chiral condensate at realistic values of the coupling 
do not see any trace of gauge field sectors except the one of topological 
charge $\nu=0$ \cite{V0}. The comparisons
one can make are therefore slightly limited. On the other hand, 
computationally the staggered fermion formulation is extremely convenient
for our purposes. We shall therefore start with a systematic study of
the chiral condensate based on staggered fermions.

For the determination of the staggered quenched chiral condensate, we used a
stochastic estimate method for several $\beta$ in SU(2) and SU(3) with
fermions in the fundamental and adjoint representations. We are
interested in solving the linear system of equations
$D^\dagger(m)D(m)\eta(m)=\left(D^\dagger(0)D(0) + m^2\right)\eta(m)=b$ for
some stochastic source $b$ and the staggered Dirac operator $D$. We
can relate the solution $\eta(m)$ in a fixed background gauge field to
the final quantities we are interested in with the following
expressions:
\begin{eqnarray}
\langle b| D^{-1}(m) |b \rangle &=& m\; b^\dagger \eta(m) \cr
\langle b| (D^\dagger(m) D(m))^{-1} + D^{-2}(m) |b\rangle &=&
   2\, m^2\; \eta^\dagger(m) \eta(m)
\label{eq:st_traces}
\end{eqnarray}
A multi-shift conjugate method~\cite{Jegerlehner} was used to solve 
the required linear system for several fermion masses.  
We supplemented these
measurements with computations of the lowest 50 eigenvalues of the
staggered Dirac operator, typically in 32 bit single precision. These
eigenvalues were used to compute a truncated spectral sum
approximation for the condensates at small fermion masses. The truncated
spectral sum method greatly improved the statistics.  However, for the
small lattices ($4^4$ and $6^4$) typically on the order of 100000
configurations were needed. As will be discussed below, this
reasonably large amount of statistics was necessary to adequately
sample the lowest eigenvalue distribution of the Dirac operator probed
by the small fermion masses used for our tests of the predictions of
Chiral Random Matrix Theory. In addition, for the orthogonal ensemble
cases, double precision was used for the eigenvalues.

\begin{figure}
\epsfxsize = 4in
\centerline{\epsffile{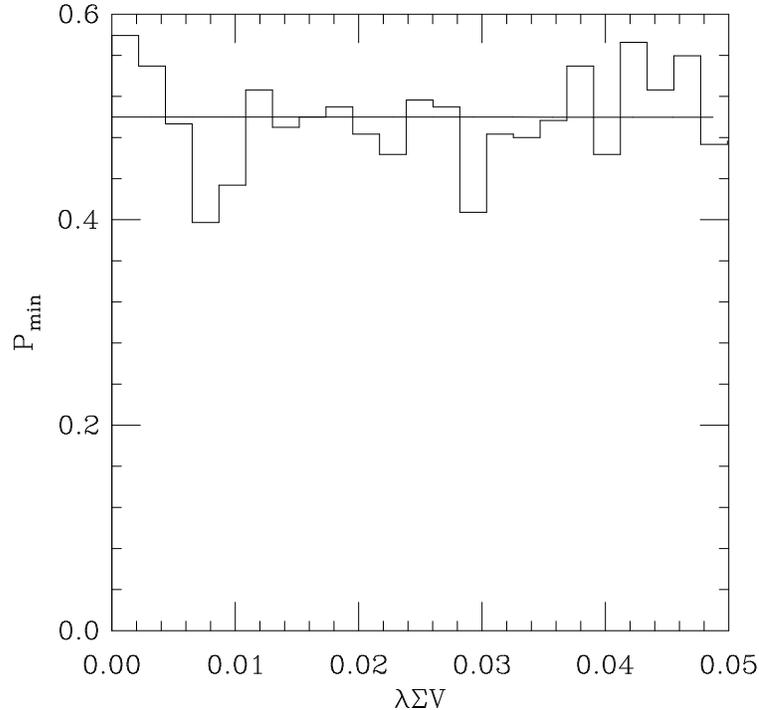}}
\caption{The distribution of the lowest eigenvalue for staggered
fermions in the adjoint representation of SU(2) at $\beta=2.0$ and
$4^4$ corresponding to the orthogonal ensemble. The scale is greatly enlarged
to show the leading edge of the distribution.}
\label{fig:pmin_orth_st}
\end{figure}

Consider first the gauge group SU(3) and staggered fermions in the
fundamental representation. The universality class is thus that of
the chUE, the same as in the continuum. In Fig.~\ref{fig:st_pbp}a we
show the quenched
chiral condensate as a function of the valence mass $m$ for a few 
different $\beta$-values, and various different
lattice sizes. At the shown values of the lattice coupling $\beta$
the infinite-volume chiral condensate $\Sigma$ has already been
determined to high accuracy from independent studies \cite{DHK}. This
means that the finite-size scaling function $\Sigma_{\nu}(\mu)$ of
Eq.~(\ref{sigmaUE}) is parameter-free. The first observation is that all the 
different data roughly collapse down on one universal scaling curve,
once plotted against $\mu=m\Sigma V$ as in Fig.~\ref{fig:st_pbp}b.
And the analytical 
prediction (\ref{sigmaUE}) for this curve is remarkably well reproduced by our 
lattice data. This of course just confirms on these lattice volumes and these
$\beta$-values the observation first made by Verbaarschot \cite{V0} on the 
basis of data from the Columbia group. It is particularly interesting
to look at the small-mass behavior. If we expand the condensate in
(\ref{sigmaUE}) for small $\mu$, we get (for $\nu=0$)
\begin{equation}
\frac{\Sigma_0^{\rm{\mbox chUE}}(\mu)}{\Sigma} = 
 - \left( \ln(\frac{\mu}{2})+\gamma - \frac{1}{2} \right) \mu
 + {\cal O}(\mu^3) ~,
\label{UEexpand}
\end{equation}
where $\gamma$ is Euler's constant.
There is the expected term linear in $\mu$, but in addition a logarithmic
correction of the form $\mu\ln(\mu)$. This latter term, which should not
be confused with so-called ``quenched chiral logarithms'', arises from
the infrared part of the integral in Eq.~(\ref{chirho}), and is thus
very sensitive to the fall-off of $\rho_s(\zeta)$ near $\zeta\sim 0$.
Indeed, for the chUE the quenched microscopic spectral density reads
\cite{V}:
\begin{equation}
\rho_s^{(0)}(\zeta) = \frac{\zeta}{2}\left[J_0(\zeta)^2 + J_1(\zeta)^2\right] ~,
\end{equation}
which for small values of $\zeta$ behaves like $\rho_s^{(0)}(\zeta) =
\zeta/2 - \zeta^3/8 + \ldots$. The leading linear term here is responsible
for the $\mu\ln(\mu)$ piece in (\ref{UEexpand}). The chUE cases with
$\nu \geq 1$ lend themselves more easily to an analysis of the low-mass 
behavior of the condensate, since the integrals
$$
\int_0^{\infty}d\zeta \frac{\rho_s^{(\nu)}(\zeta)}{\zeta^{2}}
$$
in those cases are convergent.  We can thus make the following rewriting:
\begin{equation}
\frac{\Sigma_{\nu}(\mu)}{\Sigma} ~~=~~ 
2\mu\int_0^{\infty}\!d\zeta\frac{\rho_s^{(\nu)}(\zeta)}{\zeta^2+\mu^2}~~+~~
\frac{\nu}{\mu}~~=~~
 2\mu\int_0^{\infty}\!d\zeta~\rho_s^{(\nu)}(\zeta)\left[\frac{1}{\zeta^2}
-\frac{\mu^2}{\zeta^2(\zeta^2+\mu^2)}\right]~~+~~\frac{\nu}{\mu}~,
\label{rewrite}
\end{equation}
and consider each term separately. It follows that the leading, linear, 
term in the
expansion of $\Sigma_{\nu}(\mu)$ for $\nu \geq 1$ has as coefficient the first 
Leutwyler-Smilga sum rule (as extended to this quenched case)
\cite{LS,SmV}:
\begin{equation}
\frac{\Sigma_{\nu}^{\rm{\mbox chUE}}(\mu)}{\Sigma} ~=~ 
\frac{\nu}{\mu}~+~
2\mu\left\langle \sum_{n>0}\frac{1} {\zeta_n^2}\right\rangle_{\nu} + \ldots ~=~
\frac{\nu}{\mu}~+~\frac{1}{2\nu}\mu + \ldots
\ldots
\label{eq:sumrule_UE}
\end{equation}

\begin{figure}
\centerline{{\setlength{\epsfxsize}{6in}\epsfbox[80 40 550 530]{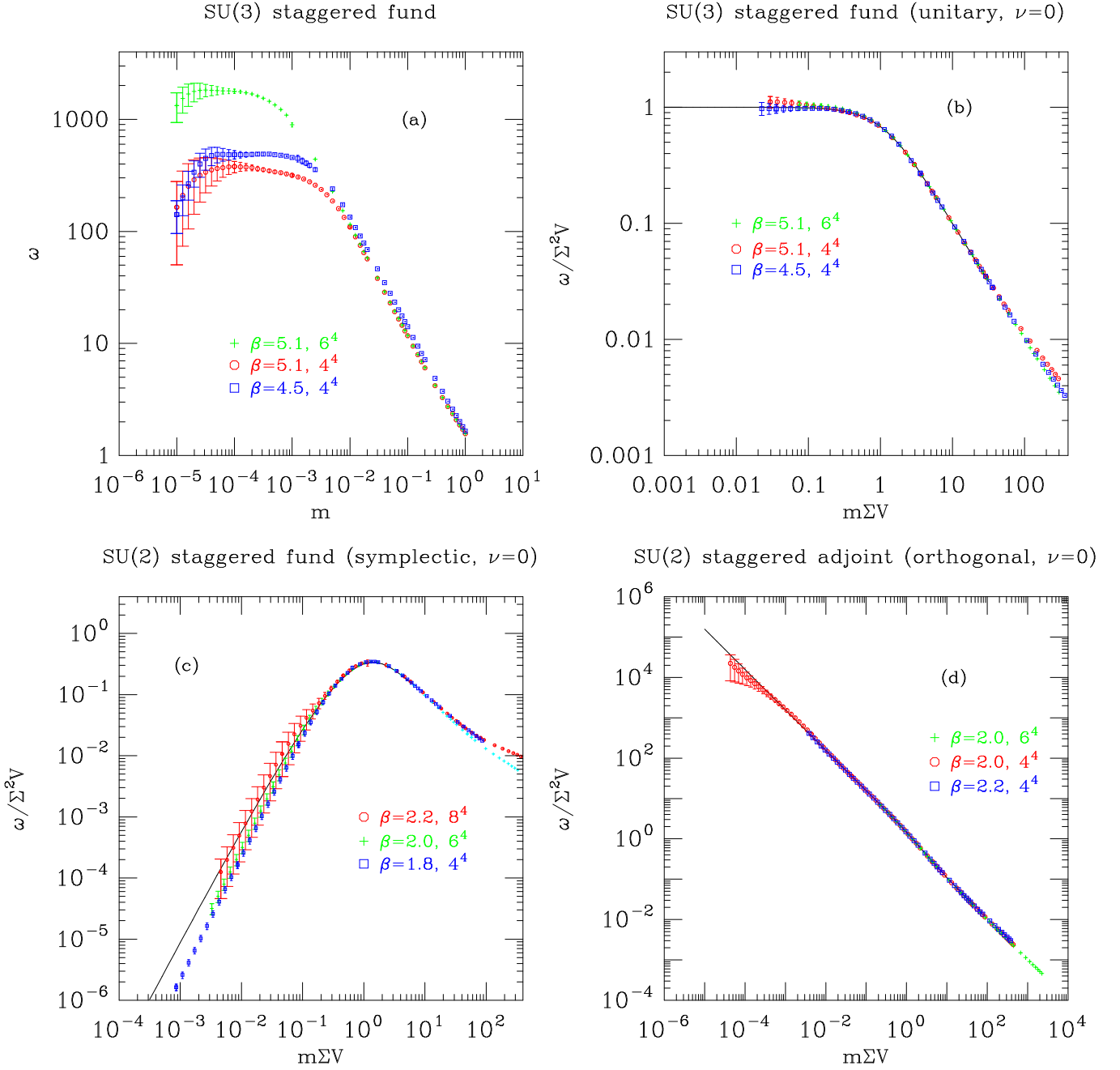}}}
\caption{The quenched susceptibility $\omega$ for staggered fermions in
(a) the fundamental representation of SU(3) as function of the fermion mass,
(b) in rescaled form as a function of $\mu = m \Sigma V$,
(c) the fundamental representation of SU(2),
(d) the adjoint representation of SU(2).}
\label{fig:st_om}
\end{figure}

We next turn to gauge group SU(2) and staggered fermions in the
fundamental representation. The analytical prediction is here that of 
the chSE universality class, with a chiral condensate as in
Eqs.~(\ref{sigmaSEeven}) and (\ref{sigmaSEodd}). 
We again choose $\beta$-values for which the 
infinite-volume chiral condensate $\Sigma$ is known to high accuracy,
so that the analytical predictions (\ref{sigmaSEeven}) and (\ref{sigmaSEodd}) 
also are parameter-free.
One remark is in order here: in the chSE universality class every eigenvalue
is doubly degenerate. The analytical predictions from RMT consider only
one of the eigenvalues from each degenerate pair. A stochastic estimate
of the condensate, on the other hand, contains the contribution from both
eigenvalues of each pair, and is thus a factor two larger. We have therefore
divided the stochastic estimate of the condensate by this factor two
in order to compare to the analytical prediction.
In Fig.~\ref{fig:st_pbp}c
we show how all data nicely collapse down on the universal scaling function
(\ref{sigmaSEeven}) for $N_f=0$. 
The agreement is seen to be extraordinarily good over
more than three orders of magnitude.

The microscopic spectral density of that case \cite{BBetal} reads as
follows for $N_f=0$:
\begin{equation}
\rho_s^{(0)}(\zeta) = \zeta J_1(2\zeta)^2  
- \frac{\pi\zeta}{2}\left(\StruveH_0(2\zeta)J_0(2\zeta)J_1(2\zeta) -
\StruveH_1(2\zeta)J_0(2\zeta)^2\right)  ~,
\end{equation}
where $\StruveH_n(x)$ is the $n$th order Struve function.
Since the small-$\zeta$ expansion is of the form $\rho_s^{(0)}(\zeta) = 
\zeta^3/3 +
\ldots$, we can make the same rewriting as above (see Eq.~(\ref{rewrite}))
and consider each term separately. It follows that also here the leading, 
linear, term in the
expansion of $\Sigma_0(\mu)$ has as coefficient the first 
Leutwyler-Smilga sum rule for that case \cite{LS,SmV}:
\begin{equation}
\frac{\Sigma_0^{\rm{\mbox chSE}}(\mu)}{\Sigma} ~=~ 
2\mu\left\langle \sum_{n>0}\frac{1}
{\zeta_n^2}\right\rangle_0 + \ldots ~=~
\mu + \ldots
\end{equation}
This linear behavior with 
coefficient one is precisely what is observed in Fig.~\ref{fig:st_pbp}c. 
The same argument goes through also for sectors of non-trivial winding
numbers $\nu$ , in which case the formula reads:
\begin{equation}
\frac{\Sigma_{\nu}^{\rm{\mbox chSE}}(\mu)}{\Sigma} ~=~ 
\frac{\nu}{\mu} + 2\mu\left\langle
\sum_{n>0}\frac{1}{\zeta_n^2}\right\rangle_{\nu} + \ldots ~=~
\frac{\nu}{\mu} + \frac{1}{1+2|\nu|}\mu + \ldots
\ldots
\label{eq:sumrule_SE}
\end{equation}

We finally present lattice gauge theory data for the SU(2) gauge group, and
staggered fermions in the adjoint representation. Here data should align
on the universal scaling curve of the chOE universality class (see
Eq.~(\ref{sigmaOE2})), and we show the results
of a few high statistics (but rather small) lattice volumes and two
different $\beta$-values 
in Fig.~\ref{fig:st_pbp}d.
The $\beta$-values were again chosen on the basis
of having already a good estimate for the infinite-volume chiral condensate
$\Sigma$ (for the adjoint representation) \cite{EHN}. The analytical curve
is seen to have a surprising behavior: it {\em rises}, even here in the
$\nu=0$ case, with 
decreasing (rescaled) fermion mass $\mu$. This unusual feature is a 
reflection of a peculiarity of the quenched microscopic spectral density
$\rho_s^{(0)}(\zeta)$ for the chOE (see the third of ref. \cite{V} and
$e.g.$ \cite{Widom}),
\begin{eqnarray}
\rho_s^{(0)}(\zeta) &=& \frac{\zeta}{2}J_1(\zeta)^2
+ \frac{1}{2}J_0(\zeta)\left[1
- \frac{1}{2}\pi\zeta\left(\StruveH_0(\zeta)J_1(\zeta) -
\StruveH_1(\zeta)J_0(\zeta)\right)\right] ~.\label{rhoOE}
\end{eqnarray}  
Contrary to all other microscopic spectral densities for the chiral
ensembles, the above function does not vanish at $\zeta=0$. This is an
artifact of the quenched limit, and it implies that in this quenched 
theory one can have spontaneous symmetry breaking even if one is taking
the limit $V\to\infty$ and $m\to 0$ in a correlated manner. Conventionally
the possibility of spontaneous symmetry breaking implies that one {\em
first} sends the volume $V$ to infinity, and only subsequently takes
the massless limit, $m\to 0$:
\begin{equation}
\Sigma ~\equiv~ \lim_{m\to 0} \lim_{V\to\infty} \langle\bar{\psi}\psi\rangle ~.
\end{equation}
In the quenched case corresponding to the chOE we observe that spontaneous
symmetry breaking can occur even if we take the simultaneous limit
$V\to \infty$ and $m\to 0$, with $mV$ fixed. This holds only in the
$\nu=0$ sector. For $\nu \neq 0$ we face the usual situation that the
chiral condensate diverges like $1/\mu$. This holds in the quenched theory
as well when we sum over topological charges \cite{top}.

Also in this case we can analyze the limit of $\mu\to 0$ analytically for
$\nu=0$. The reason for the unusual phenomenon of a constant
mass-dependent chiral condensate in the limit $\mu\to 0$ is the term 
$J_0(\zeta)/2$ in Eq.~(\ref{rhoOE}). It is this term that leads to a
non-vanishing microscopic spectral density at $\zeta=0$, and one can easily
confirm that it is also this term that is responsible for the leading
small-$\mu$ behavior of the chiral condensate in this case. Using
\begin{equation}
\int_0^{\infty}\! d\zeta \frac{J_0(\zeta)}{\zeta^2+\mu^2}
~=~ \frac{\pi}{2\mu}(I_0(\mu) - \StruveL_0(\mu))
\end{equation}
and the small-$\mu$ expansion of the modified Struve function $\StruveL_0(\mu)
= 2\mu/\pi + \ldots$, we see that only the first piece contributes in
the limit $\mu\to 0$. From Eqs.~(\ref{chirho}) and (\ref{rhoOE}) we finally 
get\footnote{Note that the $\mu\to 0$ limit gives a condensate that
is a factor of $\pi/2$ larger than the conventionally defined chiral
condensate. If finite-volume effects do not eventually cut off the lowest
eigenvalue in this case, one can even have spontaneous chiral symmetry breaking
without first taking the infinite-volume limit (we thank J. Verbaarschot
for emphasizing this last point).}.

\begin{equation}
\frac{\Sigma_0^{\rm{\mbox chOE}}(\mu)}{\Sigma} ~=~ \frac{\pi}{2} +
 ~{\cal O}(\mu)
\end{equation}
for this universality class. An approach towards this constant value
is seen in the data of Fig.~\ref{fig:st_pbp}d, but the signal
obviously gets rather noisy around $\mu \sim 10^{-3}$ for these
lattice volumes.  We show in Fig.~\ref{fig:pmin_orth_st} magnified
plot of the distribution of the smallest eigenvalue along with the
curve for the fit for the $4^4$ lattice. We see there is a reasonable
sampling of the distribution for the very small eigenvalues, but even
larger statistics beyond our 135000 configurations are needed to
really adequately sample this region and hence give very reliable
estimates for the condensate. 
Again we see that the chiral condensate is an extremely
sensitive probe of the smallest Dirac eigenvalue spectrum. For example,
the statistical fluctuation that causes a small surplus of eigenvalues
very close to the origin in Fig.~\ref{fig:pmin_orth_st} reflects itself directly in
the slightly larger chiral condensate in Fig.~\ref{fig:st_pbp}d. The
deviation is seen clearly on the non-logarithmic vertical scale.

In sectors of non-vanishing topological charge $\nu$, the 
microscopic spectral density vanishes at the origin, and if it were
not for the $\nu/\mu$-piece, the mass-dependent
chiral condensate would then also vanish as $\mu\to 0$, even in the 
infinite-volume limit. For example, for $\nu=1$ the expansion
for small $\mu$ reads in this case
\begin{equation}
\frac{\Sigma_1^{\rm{\mbox chOE}}(\mu)}{\Sigma} ~=~ 
\frac{1}{\mu} - \frac{1}{2} \left( \ln(\frac{\mu}{2}) + \gamma - 1 \right) \mu
 + {\cal O}(\mu^3) ~,
\end{equation}
with, again, a $\mu\ln(\mu)$-term in addition to the purely linear
contribution.

Other physical observables can of course be extracted from the 
finite-volume partition function. We shall here focus on one such observable,
a chiral susceptibility $\omega(\mu)$ which we define as:
\begin{equation}
\frac{\omega_{\nu}(\mu)}{\Sigma^2V} ~\equiv~ 4\mu^2 \int_0^{\infty}\! d\zeta 
\frac{\rho_s^{(\nu)}(\zeta;\mu)}{(\zeta^2+\mu^2)^2} + \frac{2\nu}{\mu^2} ~.
\label{eq:omega}
\end{equation}
We expect this quantity to be a more sensitive probe of the rescaled
eigenvalues $\zeta$ at a specific rescaled mass $\mu$ than the chiral
condensate because of the higher power occurring in the denominator of
the integrand.
This quantity is especially easy to compute in the quenched limit, where
the spectral density is $\mu$-independent. One then has
\begin{eqnarray}
\frac{\omega_{\nu}(\mu)}{\Sigma^2V}
 &=& -2\mu\frac{\partial}{\partial\mu} \left[ \int_0^{\infty} d\zeta
\frac{\rho_s^{(\nu)}(\zeta)}{(\zeta^2+\mu^2)} + \frac{\nu}{2\mu^2} \right] \cr
&=& -\mu\frac{\partial}{\partial\mu}\left.\left[\frac{\Sigma_{\nu}(\mu)}{
\mu}\right]\right/\Sigma \cr
&=& \left.\left[\frac{\Sigma_{\nu}(\mu)}{\mu} - \Sigma'_{\nu}(\mu)\right]
\right/\Sigma ~.
\end{eqnarray}
That combination is particularly useful in testing the small deviation
from linear behavior of $\Sigma_{\nu}(\mu)$ in, for instance, the case
corresponding to the chUE with $\nu=0$. In general, for a quenched 
condensate of the form
\begin{equation}
\frac{\Sigma_{\nu}(\mu)}{\Sigma} ~=~ \frac{\nu}{\mu} + A\mu +
 B\mu\ln(\mu) + C\mu^2 + \ldots
\end{equation}
we indeed find
\begin{equation}
\frac{\omega_{\nu}(\mu)}{\Sigma^2V} ~=~ \frac{2\nu}{\mu^2} - B -
 C\mu + \ldots ~.
\end{equation}
The linear term in $\Sigma_{\nu}(\mu)$  has cancelled, and the
asymptotic behavior for $\mu\to 0$ gives us the constant in front of
the $\mu\ln(\mu)$-term in $\Sigma_{\nu}(\mu)$.

For the chUE universality class, the quenched susceptibility defined above
becomes quite simple:
\begin{equation}
\frac{\omega_{\nu}^{\rm{\mbox chUE}}(\mu)}{\Sigma^2V} ~=~
2 I_{\nu+1}(\mu) K_{\nu-1}(\mu) + \frac{2\nu}{\mu^2}
\end{equation}
which has the small $\mu$ expansion
\begin{equation}
\frac{{\omega_{\nu}^{\rm{\mbox chUE}}(\mu)}}{\Sigma^2V} ~=~ 
\frac{2\nu}{\mu^2} ~+~ \cases{
1+\frac{1}{2}(\ln(\frac{\mu}{2})+\gamma-\frac{1}{4})\mu^2+{\cal O}(\mu^4)
& if $\nu=0$ \cr
-\frac{1}{4}(\ln(\frac{\mu}{2})+\gamma)\mu^2+{\cal O}(\mu^4) &
if $\nu=1$ \cr
\frac{1}{4\nu(\nu^2-1)}\mu^2+{\cal O}(\mu^4) & if $\nu\ge 2$ \cr}
\end{equation}
Of course, similar expressions can be derived for the partially quenched
cases. 

In Fig.~\ref{fig:st_om}a we show raw data for $\omega$ for the same lattice
couplings and lattice volumes as in Fig.~\ref{fig:st_pbp}. Again, these raw
data beautifully collapse down on the one single scaling function
$\omega_0(\mu)$ when rescaled according to the above prescription, as shown
in Fig.~\ref{fig:st_om}b. We emphasize again that the data for $\omega(\mu)$
are much more sensitive probes of the microscopic spectral density of the
Dirac operator than the chiral condensate itself.

For the chSE and chOE universality classes the general expressions for
$\omega(\mu)$ are quite involved, but the cases with $\nu=0$ and $1$ become
relatively simple. The prediction for the chSE universality class in 
a sector of topological charge zero and one reads:
\begin{equation}
\frac{\omega_0^{\rm{\mbox chSE}}(\mu)}{\Sigma^2V} ~=~
\pi \left[ K_0(2\mu) + 2 \mu K_1(2\mu) \right]
\left[ \StruveL_0(2\mu) I_1(2\mu) - \StruveL_1(2\mu) I_0(2\mu) \right]
 - 4 \mu K_1(2\mu) I_2(2\mu) ~,
\label{eq:o0_SE}
\end{equation}
\begin{eqnarray}
\frac{\omega_1^{\rm{\mbox chSE}}(\mu)}{\Sigma^2V} &=& \frac{2}{\mu^2} +
 4 K_1(2\mu) \left[ I_3(2\mu) + \mu I_2(2\mu) \right] +
 8 K_2(2\mu) I_2(2\mu) \cr
 && - \pi \left[ 3 K_2(2\mu) + 2\mu K_1(2\mu) \right]
\left[ \StruveL_0(2\mu) I_1(2\mu) - \StruveL_1(2\mu) I_0(2\mu) \right]
\label{eq:o1_SE}
\end{eqnarray}
\noindent
with the small $\mu$ expansion
\begin{equation}
\frac{\omega_{\nu}^{\rm{\mbox chSE}}(\mu)}{\Sigma^2V} ~=~ 
\frac{2\nu}{\mu^2} ~+~ \cases{
-\frac{4}{3}(\ln(\mu)+\gamma-\frac{1}{4})\mu^2+{\cal O}(\mu^4) &
if $\nu=0$ \cr
\frac{1}{15}\mu^2+{\cal O}(\mu^4) &
if $\nu=1$. \cr}
\end{equation}
The analogous prediction for the chOE case is
\begin{eqnarray}
\frac{\omega_0^{\rm{\mbox chOE}}(\mu)}{\Sigma^2V} &=&
\frac{\pi}{2} \left[ K_0(\mu) + \mu K_1(\mu) \right]
\left[ \StruveL_0(\mu) I_1(\mu) - \StruveL_1(\mu) I_0(\mu) \right]
 - \mu K_1(\mu) I_2(\mu) \cr
 && + \frac{\pi}{2\mu} \left[ I_0(\mu) -  \StruveL_0(\mu) \right]
 - \frac{\pi}{2} \left[ I_1(\mu) - \StruveL_1(\mu) \right] + 1 ~,
\label{eq:o0_OE}
\end{eqnarray}
\begin{equation}
\frac{\omega_1^{\rm{\mbox chOE}}(\mu)}{\Sigma^2V} ~=~ \frac{2}{\mu^2} +
 K_0(\mu) I_2(\mu) + K_1(\mu) I_1(\mu)
\label{eq:o1_OE}
\end{equation}
with the small $\mu$ expansions
\begin{equation}
\frac{\omega_{\nu}^{\rm{\mbox chOE}}(\mu)}{\Sigma^2V} ~=~ 
\frac{2\nu}{\mu^2} ~+~ \cases{
\frac{\pi}{2\mu}-\frac{\pi}{8}\mu -
\frac{1}{6}(\ln(\frac{\mu}{2})+\gamma-\frac{19}{12})\mu^2+{\cal O}(\mu^3) &
if $\nu=0$ \cr
\frac{1}{2}+\frac{1}{8}(\ln(\frac{\mu}{2})+\gamma-\frac{1}{2})\mu^2+
{\cal O}(\mu^4) &
if $\nu=1$. \cr}
\end{equation}
We show the rescaled data for gauge group SU(2) and staggered fermions
in the fundamental representation in Fig.~\ref{fig:st_om}c, and compare
these rescaled data with the analytical prediction (\ref{eq:o0_SE}).
The agreement is quite good, except for the very smallest lattice volume
($4^4$ at $\beta=1.8$).  Finally, in Fig.~\ref{fig:st_om}d we show
analogous data for gauge group SU(2) and staggered fermions in the adjoint
representation, where the analytical prediction (of the chOE universality
class) is as given in Eq.~(\ref{eq:o0_OE}). Again the agreement is perfect. 
 
From the Leutwyler-Smilga sum rules and from the small mass expansions
for the chiral condensate we can make a general prediction for the
agreement of the condensate with the RMT predictions based on how
well the microscopic spectral density fits the corresponding RMT
predictions. We see in the unitary case that the coefficient of the
linear order in the mass prediction for the sum rule
Eq.~(\ref{eq:sumrule_UE}) is dependent on all the non-zero eigenvalues
(appropriately weighted) in the spectral sum for $\nu > 0$.  For the
$\nu=0$ case, we find the condensate will depend most strongly on the
leading edge of the lowest eigenvalue for very small $\mu$. This
dependence is related to the appearance of the $\mu\ln(\mu)$ term in
Eq.~(\ref{UEexpand}).  For the symplectic case, the coefficient in
Eq.~(\ref{eq:sumrule_SE}) is predicted for all $\nu\ge 0$ (no
appearance of a $\ln(\mu)$ term at leading order). For the orthogonal
case, there is no sum-rule and we can expect a strong dependence on
the smallest non-zero eigenvalue in all topologies at small fermion
mass.  In general then if the microscopic spectral density agrees well
for many oscillations with RMT we can expect reasonable agreement for the
condensates with RMT in the ensembles and topological sectors where
the sum-rules apply. In the other cases, when probing with a small
fermion mass there can be a strong dependence on how well the smallest
eigenvalue is sampled.

At the same time, there is another competing effect that can make the
non-zero topology data fit reasonably well with the RMT predictions.
From the solution to $\eta(m)$ in Eq.~(\ref{eq:st_traces}), we see that
when $m \rightarrow 0$ the cutoff for the bottom spectrum of
$D^\dagger(0)D(0)$ is the smallest eigenvalue which will be
non-zero. Hence, $\eta$ goes to a constant at $m = 0$. However,
$\langle\bar\psi\psi\rangle$ and $\omega$ go down with an explicit $m$
or $m^2$ factor. The RMT predictions are non-trivial because, in the
unitary case for example, a $\mu\ln(\mu)$ term is generated. At higher
topology, the $\ln(\mu)$ term moves to higher ${\cal O}(\mu)$. Hence,
we can see trivial agreement with RMT at higher Q when we probe the
smallest eigenvalue.  However, we still have to get the overall
infinite volume scale $\Sigma$ correct and that is non-trivial.

\section{Topology: Overlap Fermions}

\begin{figure}
\centerline{{\setlength{\epsfxsize}{6in}\epsfbox[90 40 550 540]{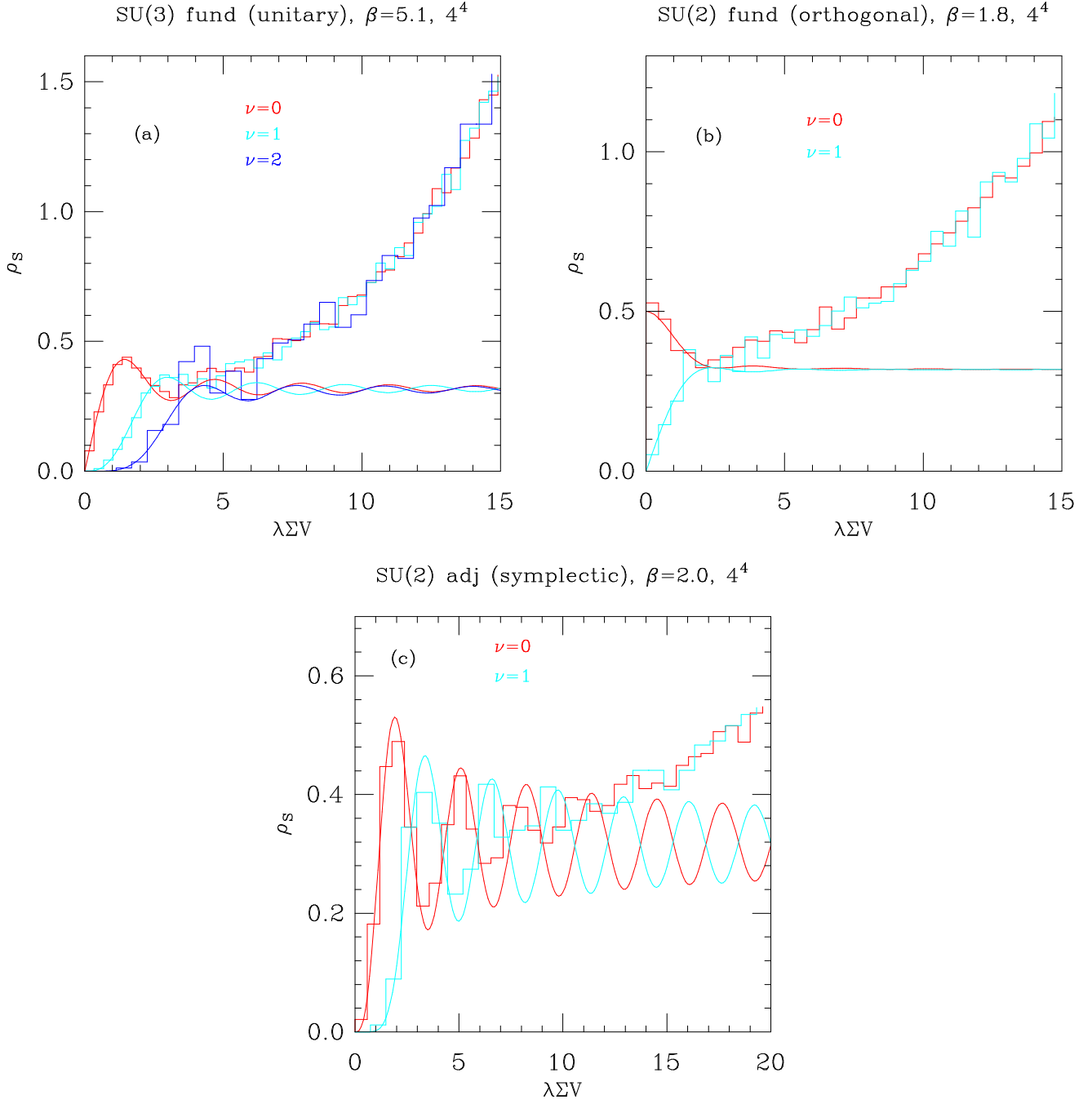}}}
\caption{
The microscopic spectral density for overlap fermions in (a)
the fundamental representation of SU(3) (unitary), 
(b) the fundamental representation of SU(2) (orthogonal), 
(c) the adjoint representation of SU(2) (symplectic).
Deviations from the analytical predictions (smooth curves) beyond a
few oscillations are due to the rather small physical volumes considered.}
\label{fig:ov_rhos}
\end{figure}

It is particularly interesting to test the analytical predictions for
sectors with non-trivial topological charge. While staggered fermions
are unsuitable for this, there now exist lattice-fermion formulations
which correctly reproduce those chiral Ward identities that are
sensitive to gauge field topology. Because they share the same Ward
identities as continuum fermions, their effective Lagrangians coincide
with those of conventional chiral perturbation theory. In particular, in
the scaling limit (1), these lattice fermions will give rise to the
same Leutwyler-Smilga effective Lagrangians (depending on the gauge groups
and color representations), and will hence fall
into exactly the same universality classes as continuum fermions.

The overlap Dirac operator \cite{herbert} derived from the overlap
formalism \cite{over} is a
proper realization of a single flavor massless fermion on the lattice
that separates lattice gauge fields into different topological classes
based on the number of exact fermion zero modes.
The massive overlap Dirac operator is given by
\begin{equation}
D(m) ~=~ 
\frac{1}{2}\left[1+m+(1-m)\gamma_5\epsilon(H_w)\right] ~. 
\label{overlapD}
\end{equation}
with $0\le m\le 1$ describing fermions with positive mass all the 
way from zero to infinity and where $H_w$ is the hermitian Wilson-Dirac
operator with a negative Wilson-Dirac mass on the lattice 
\cite{herbert,EHN1,EHN2}. Here $\epsilon(x)$ indicates the sign function.

\begin{figure}
\epsfxsize = 4in
\centerline{\epsffile{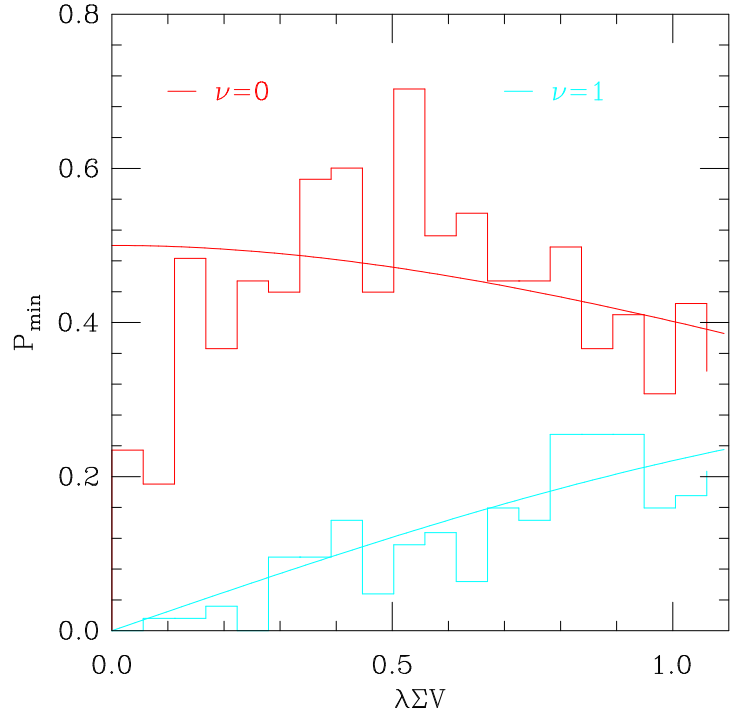}}
\caption{The distribution of the lowest eigenvalue for overlap
fermions in the orthogonal ensemble. The scale is enlarged to show the
leading edge of the distribution.}
\label{fig:pmin_orth_ov}
\end{figure}

The external fermion propagator is given by 
\begin{equation}
{\tilde D}^{-1}(m)=(1-m)^{-1}\left[D^{-1}(m) 
-1\right] ~.
\end{equation}
The subtraction at $m=0$ is evident from the original overlap
formalism~\cite{over} and the massless propagator anti-commutes with
$\gamma_5$~\cite{herbert}. With our choice of subtraction
and overall normalization the propagator satisfies the relation
\begin{equation}
m \langle b | \Bigl[ \gamma_5{\tilde D}^{-1}(m) \Bigr]^2
| b \rangle
= \langle b | {\tilde D}^{-1}(m) | b \rangle
\ \ \ \ \forall \ \ b \ \ \ {\rm satisfying} \ \ \ 
\gamma_5 | b \rangle= \pm | b\rangle
\label{eq:Goldstone}
\end{equation}
for all values of $m$ in an arbitrary gauge field background 
\cite{EHN1}.
If chiral symmetry is broken, the right hand side of Eq.~(\ref{eq:Goldstone})
is non-zero in the massless limit implying that the pion mass goes to zero
as the square root of the fermion mass.
Since $(\gamma_5 D(m))^2$ commutes with $\gamma_5$, its eigenvectors
are chiral. In the basis where $(\gamma_5 D(m))^2$ is diagonal,
$\gamma_5 D(m)$ is block diagonal with each block being a $2\times 2$ 
matrix
\cite{EHN1}. Exact zero eigenvalues of $(\gamma_5 D(m))^2$ 
are paired with
unit eigenvalues of $(\gamma_5 D(m))^2$  with the opposite chirality.
These eigenvectors of $(\gamma_5 D(m))^2$ are also eigenvectors of
$D(m)$ and therefore the topological zero modes of $D(m)$
are chiral.
We shall denote the non-zero eigenvalues of $(\gamma_5 D(m))^2$ by
$\lambda_i^2$ with $0 < \lambda_i^2 <1$. In terms of these eigenvalues, the
chiral condensate and chiral susceptibility in a fixed gauge field
background are given by \cite{EHN1}
\begin{equation}
\frac{1}{V} \sum_{x} \langle \bar\psi(x) \psi(x) \rangle_A  = 
\frac{1}{V} {\rm Tr}[{\tilde D}^{-1}]= \frac{|\nu|}{m V} +
\frac{1}{V} \sum_i \frac{2m(1-\lambda_i^2)} 
 {\lambda_i^2(1-m^2) + m^2} ~.
\label{eq:chiral}
\end{equation}
and 
\begin{equation}
\omega
= {1\over m} \langle \bar\psi \psi \rangle_A
- {d\over dm} \langle \bar\psi \psi \rangle_A
= {1\over V} \left({\rm Tr}(\gamma_5\tilde D)^{-2}(m) + 
{\rm Tr}\tilde D^{-2}(m)\right)~.
\end{equation}
Similar to Eq.~(\ref{eq:Goldstone}), we have~\cite{EHN3}
\begin{eqnarray}
\langle b | {\tilde D}^{-1}(m) | b \rangle  & = & {m \over 1-m^2} b^\dagger
\left ( \eta(m) - b \right ) \cr
\langle b | (\gamma_5\tilde D)^{-2}(m) + \tilde D^{-2}(m) | b \rangle 
& = & {2m^2\over (1-m^2)^2} \left(\eta^\dagger(m) - b^\dagger\right)
\left ( \eta(m) - b \right ) 
\label{eq:simple}
\end{eqnarray}
where
\begin{equation}
H_o^2(m)\eta(m) = b \qquad {\rm with} \qquad H_o(m) = \gamma_5 D(m),\quad \gamma_5 b = \pm b \quad.
\end{equation}
We note that 
\begin{equation}
H_o^2(m) = D^\dagger(m) D(m) = D(m) D^\dagger(m) = 
\left(1-m^2 \right) \left[ H_o^2(0) + {m^2\over 1-m^2} \right]
\label{eq:H_o_m}
\end{equation}
with
\begin{equation}
H_o^2(0)\chi_\pm = \left[{1\over 2} + {1\over 4} \left(\gamma_5\pm
1)\right)\epsilon(H_w)\right]\chi_\pm \qquad {\rm with} \quad \gamma_5
\chi_\pm = \pm \chi_\pm \quad.
\label{eq:H_o_0}
\end{equation}
Eq.~(\ref{eq:H_o_m}) implies that we can solve the set of equations
$H_o^2(m) \eta(m) = b$ for several masses, $m$, simultaneously
(for the same right hand $b$) using
the multiple Krylov space solver described in
Ref.~\cite{Jegerlehner}. In our tests we used fermion masses from $0$ to
$0.999$. However, for comparisons to RMT we only consider the
fermion mass range from $m=10^{-4}$ up to $0.999$.

The first term on the right hand side of Eq.~(\ref{eq:chiral}) is due
to the presence of $|\nu|$ exact zero modes in a fixed gauge field
background.  By working in the chiral sector where $(\gamma_5 D(m))^2$
has no zero modes, it is possible to compute the second term in
Eq.~(\ref{eq:chiral}) and investigate the onset of chiral symmetry
breaking on the lattice \cite{EHN1}. Note that the relation
(\ref{eq:chiral}) is an exact identity at any lattice spacing. It is
of the same form as Eq.~(\ref{chirho}), up to terms vanishing with the
ultraviolet cut-off.  The bare fermion mass enters the overlap Dirac
operator in a non-trivial way and is proportional to the mass
parameter $m$ in (\ref{overlapD}) only for small $m$, {\it i.e.} only
up to terms of relative ${\cal O}(a^2)$.  The proportionality factor,
$Z_m$, depends in particular on the mass in the Wilson-Dirac operator
used~\cite{EHN1}. Since $Z_m$ is the inverse of the wave function
renormalization constant~\cite{EHN1}, the rescaled mass parameter $\mu
= m\Sigma V$ is independent of these $Z$-factors and agrees with the
continuum definition up to terms vanishing with the ultraviolet
cut-off.

\begin{figure}
\centerline{{\setlength{\epsfxsize}{6in}\epsfbox[90 40 550 540]{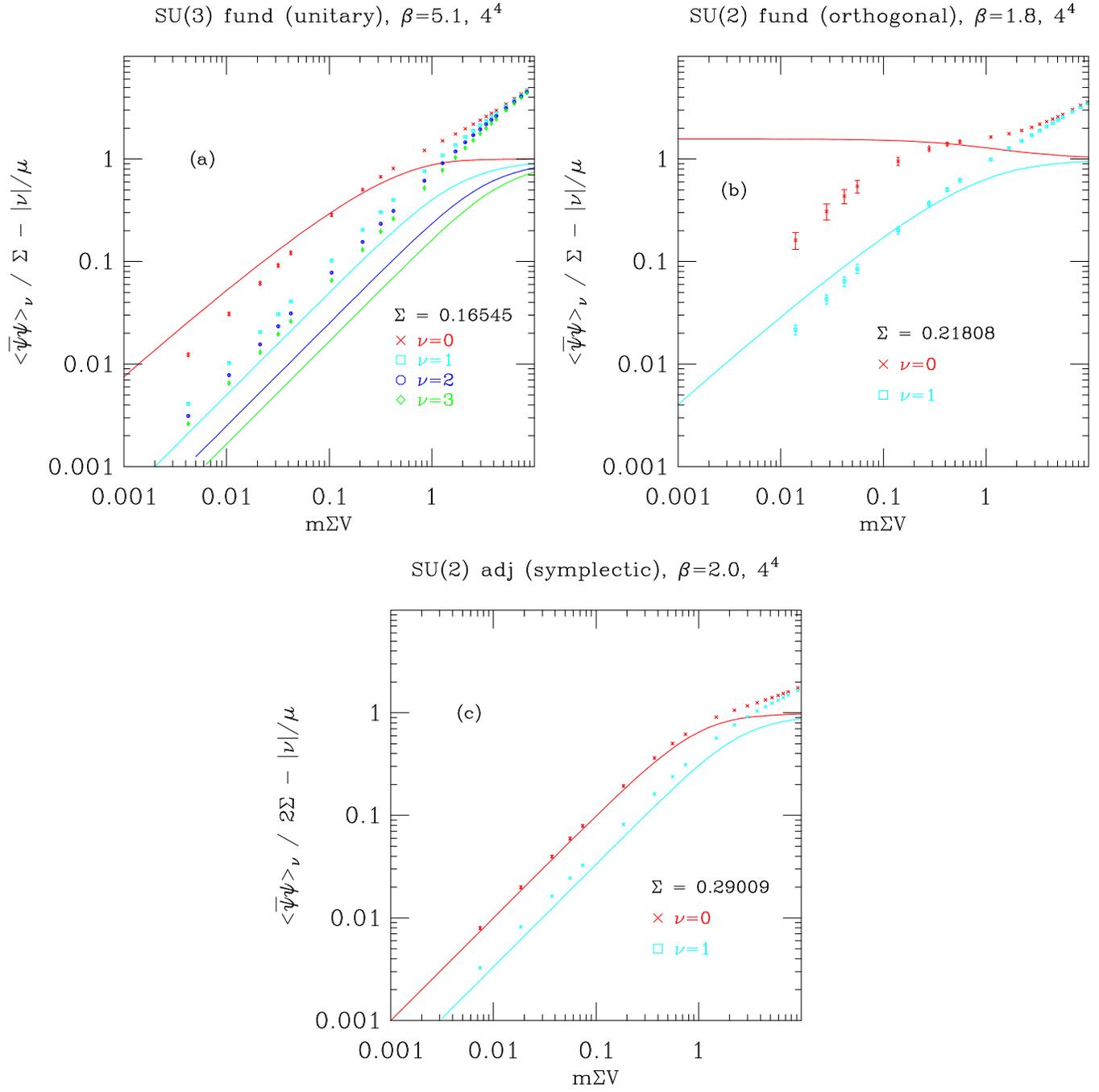}}}
\caption{The rescaled condensate for overlap fermions in (a) the fundamental
representation of SU(3) as function of $\mu = m \Sigma V$,
(b) the fundamental
representation of SU(2),
(c) the adjoint
representation of SU(2).}
\label{fig:ov_pbp}
\end{figure}

The infinite-volume
chiral condensate $\Sigma$ differs significantly, at the same $\beta$-values,
from that of staggered fermions. However, in the cases we shall present here
this one single parameter $\Sigma$ has already been extracted to high
precision from the distribution of the smallest Dirac operator eigenvalue
\cite{EHKN}. The analytical predictions for $\Sigma_{\nu}(\mu)$ are
thus parameter-free also in these cases.

On all the gauge configurations used for the measurement of the condensate
a few low lying eigenvalues have previously been determined~\cite{EHKN}.
We thus know the number and chirality of all zero modes, and hence the
topological charge. As mentioned already, when zero modes are present,
we perform the stochastic estimate in the sector with opposite chirality.
In topologically trivial gauge fields, we perform the stochastic estimate
in the positive chirality sector.

We are now ready to test some of the predictions for overlap fermions
in the finite-volume regime.
The first observation is that the universality
classes of continuum fermions coincide with those of overlap
fermions. We begin by comparing the microscopic spectral density
$\rho_s(\zeta)$ with
the predictions of RMT.
In Fig.~\ref{fig:ov_rhos} are the results for different topological
sectors for all the ensembles. The curves are the predictions using
the infinite volume $\Sigma$ previously determined~\cite{EHKN}. We see
good agreement for the first oscillation (essentially the lowest
eigenvalue contribution) for all topological sectors and the best
agreement in the symplectic case in Fig.~\ref{fig:ov_rhos}c. However, the
data rapidly deviates up from the curve for higher eigenvalues. This
is a finite-volume effect since as the volume increases the scale of the
eigenvalues decreases into the region where RMT applies. 
The infinite volume chiral condensate $\Sigma$ sets the scale for
the eigenvalues, and since the condensate is larger for staggered
fermions compared to overlap fermions at corresponding parameters, the
eigenvalues for staggered fermions occur closer to zero where there is a
corresponding better agreement for more oscillations with the RMT
predictions. 

Nevertheless, we can expect to find the best agreement of $\omega_\nu$
with the RMT predictions where the fermion mass is probing the scale
of eigenvalues in $\rho_s(\zeta)$ that are in best agreement.
For the chiral condensate $\langle\bar\psi\psi\rangle_\nu$ the
agreement with RMT will depend on the overall $\rho_s(\zeta)$ which we
remarked is best for the symplectic case.  Since the spectral density
does not match the RMT prediction for many oscillations in the unitary
case as seen in Fig.~\ref{fig:ov_rhos}a, we expect higher topologies
to not be well described by RMT.

In addition, we can expect problems for comparisons of
observables for small fermions masses when the $\rho_s(z)$ has not
been adequately sampled, and this problem is most pronounced in
the orthogonal case. Compared to the staggered fermion example in
Fig.~\ref{fig:pmin_orth_st}, we do not have enough statistics and
large enough volumes to adequately sample the leading edge of the
eigenvalue distributions shown in detail in
Fig.~\ref{fig:pmin_orth_ov}.

We continue with tests of the predictions for $\Sigma_\nu(\mu)$ in the
chUE case, using quenched overlap fermions and gauge group SU(3).
Shown in Fig.~\ref{fig:ov_pbp}a are some data for gauge field sectors
with $\nu= 0\ldots 3$. We stress that for the sectors of non-vanishing
$\nu$ we have subtracted the somewhat trivial $\nu/\mu$ term, which
otherwise would completely dominate the plot. What is shown is thus
not the chiral condensate per se, but rather
$\Sigma_{\nu}(\mu)/\Sigma-|\nu|/\mu$. The agreement in the $\nu=0$
sector is good, but while the data for the $\nu=1$, 2, and 3 qualitatively
display the right behavior, they are nevertheless somewhat off the
analytical predictions.

\begin{figure}
\centerline{{\setlength{\epsfxsize}{6in}\epsfbox[90 40 550 540]{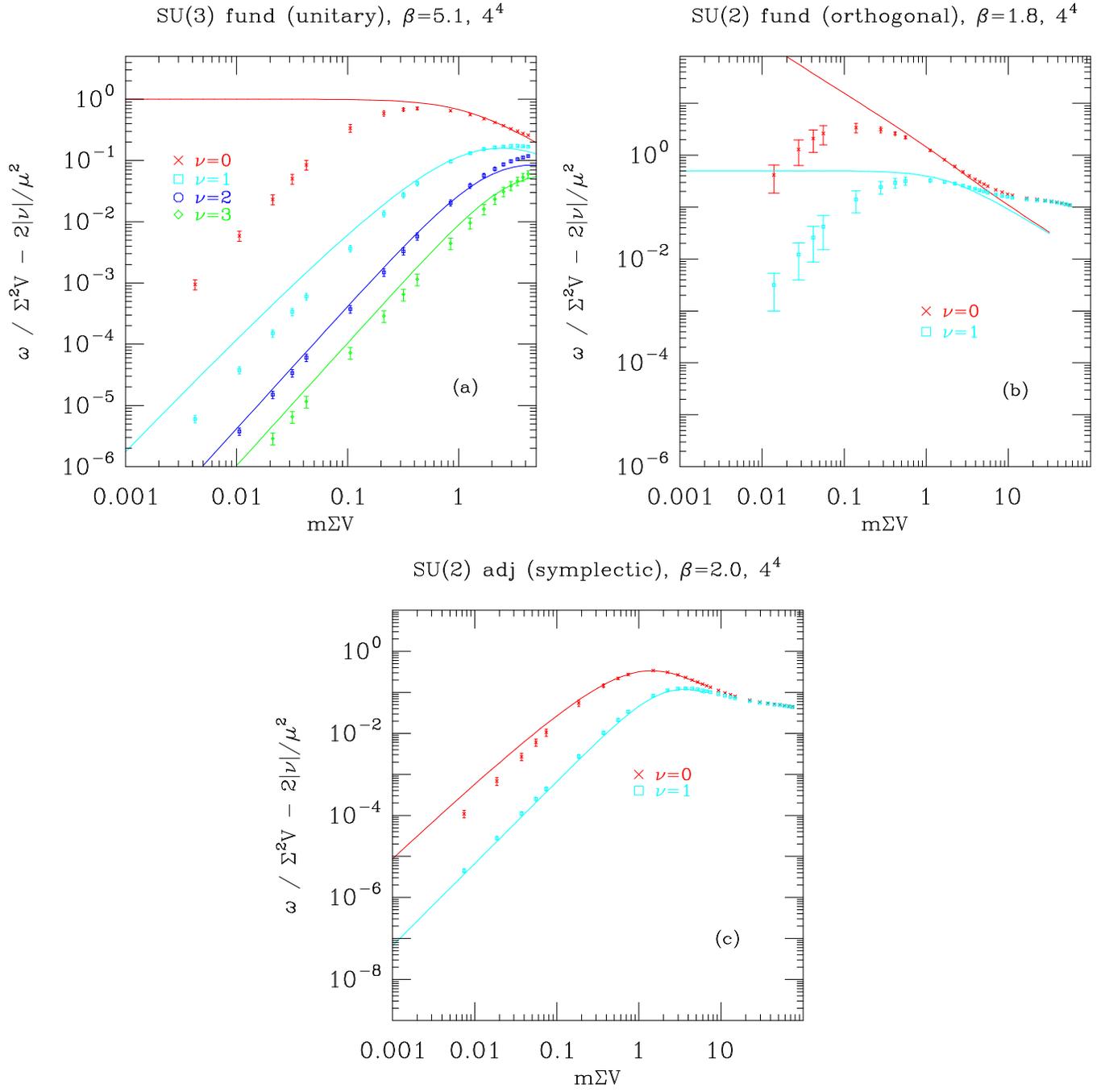}}}
\caption{The rescaled quenched susceptibility $\omega$ for overlap
fermions in
(a) the fundamental representation of SU(3) as a function of $\mu = m \Sigma V$.
(b) the fundamental representation of SU(2),
(c) in the adjoint representation of SU(2).}
\label{fig:ov_om}
\end{figure}

In Fig.~\ref{fig:ov_om}a we show the subtracted $\omega_\nu/\Sigma^2 V$
for the unitary case. As mentioned before, this observable is a more
sensitive probe of the eigenvalue distribution at a fixed fermion mass
and should yield better agreement when the corresponding
$\rho_s(\zeta)$ is in good agreement with the RMT predictions.  We see
good agreement for $z=m\Sigma V > 0.05$ in all topology sectors, but
below this value there are deviations related to the lack of small
eigenvalues in $\rho_s$ and the small volume.  For the orthogonal
ensemble we find even worse agreement for $\Sigma_{\nu}(\mu)/\Sigma -
|\nu|/\mu$ shown in Fig.~\ref{fig:ov_pbp}b, but find a small mass
region ($0.4\ltapprox z\ltapprox 4$) in $\omega_\nu/\Sigma^2 V$ shown in
Fig.~\ref{fig:ov_om}b where we simultaneously have an adequately
sampled distribution of very small eigenvalues and good agreement of
$\rho_s(z)$ with RMT.

We next turn to gauge group SU(2) and overlap fermions in the adjoint
representation where we find the expected good agreement with analytical
predictions for the $\nu=0$ and $1$ sectors (we found almost no
configurations in the sectors of higher topological charge in this
case). These graphs are shown in Fig.~\ref{fig:ov_pbp}c and
Fig.~\ref{fig:ov_om}c.

\section{Conclusions}

We have performed a systematic series of Monte Carlo tests of the
analytical predictions for the chiral condensate and related chiral
susceptibilities in the finite-volume scaling region of Eq.~(1).
In four dimensions there are three universality classes with which
to compare, conveniently classified in Random Matrix Theory terminology
by means of chiral versions of the three classical matrix ensembles, $i.e.$,
chSE, chUE and chOE. Once the infinite-volume chiral condensate $\Sigma$
is known, there are parameter-free finite-volume scaling functions
with which to compare data. As we have shown, results for all
three universality classes with topological charge $\nu=0$ are nicely 
reproduced by staggered fermions. To test the analytical predictions
for gauge field sectors of non-trivial topological winding numbers,
we have also considered overlap fermions, which possess exact zero modes
in topologically non-trivial gauge fields. Here there is qualitatively
good agreement, with even excellent
agreement in the case of the chSE universality class.
The deviations from the RMT predictions observed in the other cases
are understood to be due to either finite volume effects or the
relatively modest statistics we were able to obtain in the simulations
with overlap fermions.

The results presented here clearly show the power of the finite-size
analysis that has come out of the study of finite-volume effective partition
functions and Random Matrix Theory. In contrast to conventional finite-size
scaling analysis near critical points, we are here in the unusual situation 
of knowing not only the right scaling variables, but also parameter-free 
exact analytical predictions for the scaling quantities.
In this particular corner of those non-Abelian or Abelian gauge theories 
that support spontaneous breaking of chiral symmetry the
exact analytical predictions have thus very clearly been confirmed by direct 
numerical studies.

\vspace{1cm}

{\sc Acknowledgments:} The work of P.H.D. has been partially supported
by EU TMR grant no. ERBFMRXCT97-0122, and the work of R.G.E. and U.M.H.
has been supported in part by DOE contracts DE-FG05-85ER250000 and
DE-FG05-96ER40979. In addition, P.H.D. and U.M.H. acknowledge the financial
support of NATO Science Collaborative Research Grant no. CRG 971487.
This work was completed at the Aspen Center for Physics,
and we thank the center for the excellent working conditions provided to us.

\end{document}